\documentstyle{cupconf}
\input epsf
\ifoldfss
\else
  \ifnfssone
    \newmathalphabet{\mathit}
      \addtoversion{normal}{\mathit}{cmr}{m}{it}
      \addtoversion{bold}{\mathit}{cmr}{bx}{it}
    \newmathalphabet{\mathcal}
      \addtoversion{normal}{\mathcal}{cmsy}{m}{n}
    \else
    \ifnfsstwo
    \fi
  \fi
\fi
\def\deg{$^{\small \circ}\,$}
\def\etal{\mbox{\it et al.}}
\title[Bars and boxy/peanut-shaped bulges]{Bars and boxy/peanut-shaped bulges:\\
  an observational point of view}
\author[M. Bureau {\it et al.\/}]{M\ls A\ls R\ls T\ls I\ls N\ns B\ls U\ls R\ls
  E\ls A\ls U$^1$,\ns K.\ns C.\ns F\ls R\ls E\ls E\ls M\ls A\ls N$^2$\ns\\
  \and\ns E.\ns A\ls T\ls H\ls A\ls N\ls A\ls S\ls S\ls O\ls U\ls L\ls A$^3$}
\affiliation{$^1$Sterrewacht Leiden, Postbus~9513, 2300~RA Leiden, The
  Netherlands\\[\affilskip] $^2$Research School of Astronomy and Astrophysics,
  Institute of Advanced Studies,\\ The Australian National University, Mount
  Stromlo Observatory,\\ Private Bag, Weston Creek P.O., ACT~2611,
  Australia\\[\affilskip] $^3$Observatoire de Marseille, 2~place Le Verrier,
  F-13248 Marseille Cedex~4, France}
\begin{document}
\ifnfssone
\else
  \ifnfsstwo
  \else
    \ifoldfss
      \let\mathcal\cal
      \let\mathrm\rm
      \let\mathsf\sf
    \fi
  \fi
\fi
\maketitle
%
%
\begin{abstract}
  Prompted by work on the buckling instability in barred spiral galaxies, much
  effort has been devoted lately to the study of boxy/peanut-shaped (B/PS)
  bulges. Here, we present new bar diagnostics for edge-on spiral galaxies
  based on periodic orbits calculations and hydrodynamical simulations. Both
  approaches provide reliable ways to identify bars and their orientations in
  edge-on systems. We also present the results of an observational search for
  bars in a large sample of edge-on spirals with and without B/PS bulges. We
  show that most B/PS bulges are due to the presence of a thick bar viewed
  edge-on while only a few may be due to accretion. This strongly supports the
  bar-buckling mechanism for the formation of B/PS bulges.
\end{abstract}
%
%
\firstsection
\section{Introduction}
\label{sec:intro}
Boxy/peanut-shaped (B/PS) bulges have, as their name indicates, excess light
above the plane. They are thus easily identified in edge-on systems and
display many interesting properties: their luminosity excess, an extreme
three-dimensional structure, probable cylindrical rotation, etc.  However, the
main importance of B/PS bulges resides in their incidence: at least 20-30\% of
all spiral galaxies possess a B/PS bulge. They are thus essential to our
understanding of bulge formation and evolution.

Early theories on the formation of B/PS bulges were centered around accretion
scenarios, where one or many satellites galaxies are accreted onto a
preexisting bulge, and which lead to axisymmetric structures (e.g.\ Binney \&
Petrou 1985). However, such scenarios are restrictive, and it seems that the
only viable path is the accretion of a small number of moderate-sized
satellites. Thus, accretion probably plays only a minor role in the formation
of B/PS bulges. A more attractive mechanism is the buckling of a bar, due to
vertical instabilities. This process can form B/PS bulges even in isolated
galaxies, and accounts easily for the fact that the fraction of B/PS bulges is
similar to that of (strongly) barred spirals. Soon after a bar is formed, it
buckles and settles with an increased thickness, appearing boxy or
peanut-shaped depending on the viewing angle (e.g.\ Combes \etal\/ 1990).
Hybrid scenarios, where a bar is excited by an interaction and then buckles,
have also been suggested.

To test as directly as possible the bar-buckling hypothesis, we have developed
reliable bar diagnostics for edge-on spirals (Bureau \& Athanassoula 1999;
Athanassoula \& Bureau 1999), and have searched for bars in a sample of
edge-on galaxies with and without B/PS bulges (Bureau \& Freeman 1999). This
way, we can probe the exact relationship between bars and B/PS bulges.
%
%
\section{Bar diagnostics in edge-on spiral galaxies: the periodic orbits
  approach}
\label{sec:orbits}
There is no reliable photometric way to identify a bar in an edge-on spiral
galaxy. However, \cite{km95} showed that an edge-on barred disk produces
characteristic double-peaked line-of-sight velocity distributions which would
not occur in an axisymmetric disk. Following their work, we also developed bar
diagnostics based on the position-velocity diagrams (PVDs) of edge-on disks,
which show the projected density of material as a function of line-of-sight
velocity and projected position. The mass model we adopted has a Ferrers bar,
two axisymmetric components yielding a flat rotation curve, and four free
parameters. All our models are two-dimensional.

%
%
\begin{figure}
  \epsfysize=1.9in
  \centerline{\epsfbox{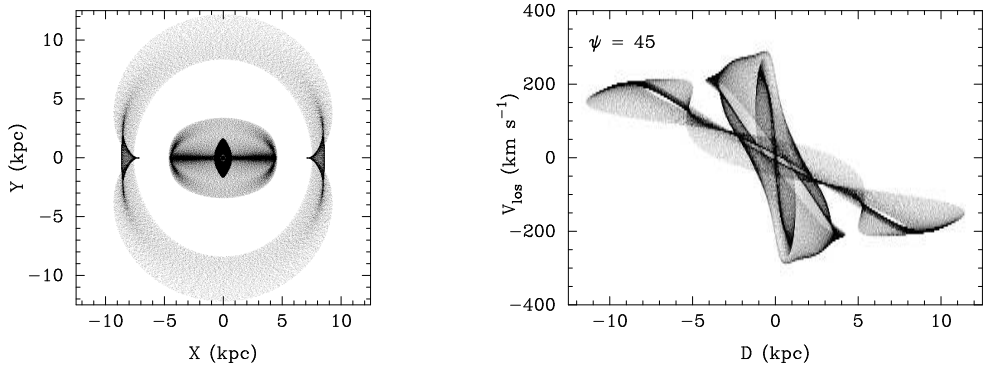}}
  \caption{Surface density distribution ({\it left}) and projected PVD ({\it
      right}) of model 001, when considering the $x_1$, $x_2$, and outer 2:1
    families of periodic orbits. In the left plot, the bar is horizontal, has
    a length of 5~kpc, and is viewed at an angle of 45\deg for the PVD.}
  \label{fig:orbits}
\end{figure}
We first used the families of periodic orbits in our mass model as building
blocks to model real galaxies (Bureau \& Athanassoula 1999). Such an approach
provides essential insight into the (projected) kinematics of spirals. We
showed that the global appearance of a PVD can be used as a reliable tool to
identify bars in edge-on disks. Specifically, the presence of gaps between the
signatures of the various periodic orbit families follows directly from the
non-homogeneous distribution of orbits in a barred galaxy. The two so-called
forbidden quadrants of the PVDs are also populated because of the elongated
shape of the orbits. Figure~\ref{fig:orbits} shows the surface density and
projected PVD of a typical model. The bar is viewed at an angle of 45\deg from
the major axis and only the major families of periodic orbits are considered.
The signatures of the $x_1$ (parallel to the bar) and $x_2$ (perpendicular to
the bar) orbits are particularly important to identify the bar and constrain
the viewing angle. Because of streaming, the parallelogram-shaped signature of
the $x_1$ orbits reaches very high radial velocities when the bar is seen
end-on and only relatively low velocities when it is seen side-on. The
opposite is true for the $x_2$ orbits.
%
%
\section{Bar diagnostics in edge-on spiral galaxies: hydrodynamical
  simulations}
\label{sec:hydro}
We also developed bar diagnostics using hydrodynamical simulations, targeting
specifically the gaseous component of spiral galaxies (Athanassoula \& Bureau
1999). The simulations are time-dependent and the gas is treated as ideal,
isothermal, and non-viscous. We used the same mass model as above, without
self-gravity, and modeled star formation and mass loss in a simplistic way.
However, the collisional nature of the gas leads to better bar diagnostics
than the periodic orbits approach.

%
%
\begin{figure}
  \epsfysize=1.9in
  \centerline{\epsfbox{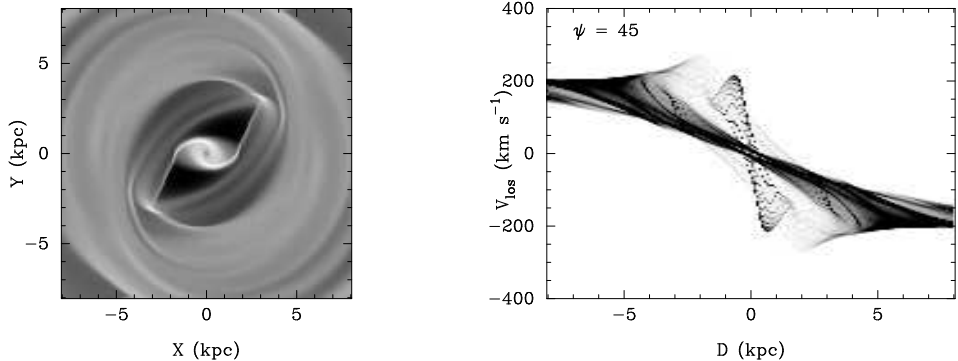}}
  \caption{Surface density distribution ({\it left}) and projected PVD ({\it
      right}) of model 001, for the hydrodynamical simulations. In the left
    plot, the bar is now oriented diagonally and high densities are in white.}
  \label{fig:hydro}
\end{figure}
The main feature of the PVDs is a gap, present at all viewing angles, between
the signature of the nuclear spiral (associated with $x_2$ orbits) and that of
the outer parts of the disks. There is very little gas in $x_1$-like flows.
This gap unmistakably reveals the presence of a bar in an edge-on disk. It
occurs because the large scale shocks which develop in bars drive an inflow of
gas toward the centers, depleting the outer bar regions. If a galaxy has no
inner Lindblad resonance (ILR; or, equivalently, has no $x_2$ orbits), there
is no nuclear spiral and the entire bar region is depleted. Then, the use of
stellar kinematics is probably preferable to identify a bar. We will develop
such diagnostics in a future paper. Figure~\ref{fig:hydro} shows the gas
density distribution and PVD for the same model as above, which has ILRs.
Although not shown, the PVDs again vary significantly with the viewing angle,
the signature of the nuclear spiral reaching its highest velocities when the
bar is seen close to side-on. We also ran simulations covering a large
fraction of the parameter space likely to be occupied by real galaxies. The
PVDs can then be used to somewhat constrain the mass distribution and bar
properties of observed systems.
%
%
\section{The nature of boxy/peanut-shape bulges}
\label{sec:obs}
The PVDs produced are directly comparable to kinematic observations of edge-on
spiral galaxies. In the hope of understanding the formation mechanism of B/PS
bulges, we searched for bars in a sample of 30 edge-on spirals with and
without B/PS bulges, using emission line long-slit spectroscopy (Bureau \&
Freeman 1999). The objects were selected from existing catalogs and 2/3 have
probable companions. Of the 24 galaxies with a B/PS bulge, 17 have extended
emission lines and constitutes our main sample. The remaining 6 galaxies all
have extended emission and form a control sample.

%
%
\begin{figure}
  \epsfysize=1.9in
  \centerline{\epsfbox{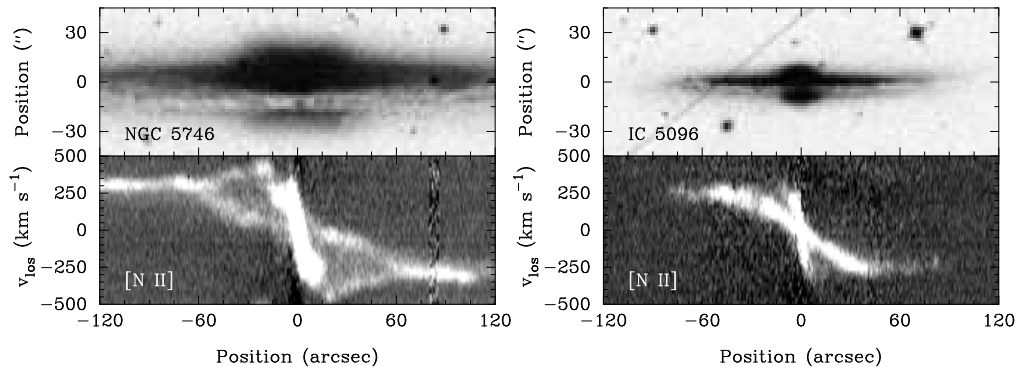}}
  \caption{Image and ionised gas PVD (on the same scale and along the major
    axis) of two B/PS galaxies. The bar signature in the bulge region is
    obvious in both cases.}
  \label{fig:obs}
\end{figure}
In the main sample, 14 galaxies display a clear bar signature in their PVD,
and only 3 may be axisymmetric or have suffered interactions. None of the
galaxies in the control sample shows evidence for a bar. This means that most
B/PS bulges are due to the presence of a thick bar viewed edge-on and only a
few may be due to the accretion of external material. In addition, spheroidal
bulges do appear axisymmetric. Thus, it seems that most B/PS bulges are
edge-on bars and that most bars are B/PS when viewed edge-on. However, the
strength of this converse is limited by the small size of the control sample.
To illustrate our data, we show the PVD of two galaxies in the main sample in
figure~\ref{fig:obs}. Our association of bars and B/PS bulges is supported by
the anomalous emission line ratios observed in many objects. These galaxies
display large H$\alpha$/[N~II] ratios, often associated with shocks, and these
ratios correlate with kinematical structures in the disks. Constraining the
viewing angle to the galaxies with our models, the observations also appear to
confirm the general prediction of $N$-body simulations, that bars are
peanut-shaped when seen side-on and boxy-shaped when seen end-on.

Our results are consistent with the current knowledge on the bulge of the
Milky Way and strongly support the bar-buckling mechanism for the formation of
B/PS bulges. However, we do not test directly for buckling, and other
bar-thickening mechanisms and hybrid scenarios can not be excluded.
Nevertheless, it is clear that the influence of bars on the formation and
evolution of bulges is primordial.
%
%
\section{On-going studies}
\label{sec:ongoing}
The bar diagnostics we have developed open up for the first time the
possibility of studying the vertical structure of bars observationally. To
this end, we have obtained $K$-band images of all the sample galaxies. We have
also obtained absorption line spectroscopic data to study the stellar
kinematics, and a more in-depth investigation of line ratios will give us a
better understanding of the large scale effects of bars in disks.
%
%
\begin{acknowledgments}
  We would like to thank A.\ Bosma, A.\ Kalnajs, and L.\ Sparke for useful
  discussions at various stages of this work. We also thank J.-C.\ Lambert for
  computer assistance and G.\ D.\ Van Albada for the FS2 code.
\end{acknowledgments}
%
%

%
\end{document}